\documentclass{Interspeech2024}
\usepackage{multirow}
\usepackage{tabularx}
\usepackage{hyperref}
\usepackage{subfigure}

\interspeechcameraready

\title{FA-GAN: Artifacts-free and Phase-aware High-fidelity GAN-based Vocoder}

\name[affiliation={2}]{Rubing}{Shen}
\name[affiliation={1,2,\dagger}]{Yanzhen}{Ren}
\name[affiliation={2}]{Zongkun}{Sun}

\address{
  $^1$Key Laboratory of Aerospace Information Security and Trusted Computing, Ministry of Education\\
  $^2$School of Cyber Science and Engineering, Wuhan University}
\email{\thanks{$\dagger$: corresponding author.}rbshen@whu.edu.cn, renyz@whu.edu.cn, zongksun@whu.edu.cn}


\begin{document}

\maketitle

\begin{abstract}
Generative adversarial network (GAN) based vocoders have achieved significant attention in speech synthesis with high quality and fast inference speed. However, there still exist many noticeable spectral artifacts, resulting in the quality decline of synthesized speech. In this work, we adopt a novel GAN-based vocoder designed for few artifacts and high fidelity, called FA-GAN. To suppress the aliasing artifacts caused by non-ideal upsampling layers in high-frequency components, we introduce the anti-aliased twin deconvolution module in the generator. To alleviate blurring artifacts and enrich the reconstruction of spectral details, we propose a novel fine-grained multi-resolution real and imaginary loss to assist in the modeling of phase information. Experimental results reveal that FA-GAN outperforms the compared approaches in promoting audio quality and alleviating spectral artifacts, and exhibits superior performance when applied to unseen speaker scenarios.
\end{abstract}

\section{Introduction}
The success of deep generative models has significantly advanced the field of speech synthesis, making it possible to convert intermediate acoustic features into natural and intelligible speech. 
The evolution of deep generative models encompasses autoregressive models \cite{oord2016wavenet,mehri2016samplernn}, flow-based models~\cite{prenger2019waveglow,oord2018parallel,lee2020nanoflow}, generative adversarial network (GAN) based models~\cite{kumar2019melgan,kong2020hifi,yamamoto2020parallel,jang2021univnet,huang2022singgan,lee2022bigvgan,kaneko23_istftnet2,dang23b_lightvoc} and diffusion models~\cite{chen2020wavegrad,lee2021priorgrad,koizumi22_specgrad}.
Among them, GAN-based vocoders have garnered widespread attention for their ability to generate high-fidelity speech. 
Although recent GAN-based vocoders synthesize almost realistic audio, there still exists a gap between the ground truth and generated audio samples in the frequency domain~\cite{kim21f_fre,pons2021upsampling,bak2023avocodo,shang2023analysis}.
It can be primarily attributed to the following aspects: 
1) aliasing artifacts, which arise from the up-sampling operations when increasing resolution. 
2) blurring artifacts, caused by the lack of phase information and spectral details in the frequency domain. 

Aliasing artifacts are the typical artifacts in GAN-based vocoders for that the modeling of high-frequency is dependent on upsampling layers to increase the input of low-dimensional features up to high-dimensional waveforms. 
Especially, transposed convolution layers are typically used to obtain high-frequency components from the low-frequency spectrograms. 
However, as~\cite{pons2021upsampling,karras2021alias} revealed, aliasing artifacts are observed in the high-frequency areas, leading to the decline of synthesized quality. 
To address this problem, 
recent works have endeavored to minimize spectral artifacts by developing enhanced discriminators or introducing additional processes alongside the original transposed convolution~\cite{bak2023avocodo,lee2022bigvgan,shang2023analysis}. 
Improved structures of discriminators are proposed to reduce artifacts, such as the collaborative multi-band discriminator (CoMBD) and sub-band discriminator (SBD)~\cite{bak2023avocodo}. 
Moreover, the low-pass filter is adopted in the generator to eliminate unwanted high-frequency components~\cite{lee2022bigvgan, shang2023analysis}.
However, the approaches mentioned above primarily focus on addressing the issues associated with transposed convolution layers by enhancing discriminative capabilities or adding extra components, yet their effectiveness in mitigation is somewhat limited.

Except for the GAN-specific aliasing artifacts in high-frequency areas, generated mel-spectrograms still suffer from blurring artifacts and a lack of explicit harmonic details.
Many discriminators and auxiliary losses are proposed to promote the modeling ability of the generator and sharpen the generated spectrograms. 
The multi-scale discriminator (MSD) is designed to analyze waveforms across different scales~\cite{kumar2019melgan}, and the multi-period discriminator (MPD) focuses on modeling periodic patterns of audio signals~\cite{kong2020hifi}. 
Additionally, the multi-resolution discriminator (MRD) has been introduced to process spectrograms at multiple resolutions, thereby enriching the spectral complexity of the synthesized waveforms~\cite{yamamoto2020parallel,jang2021univnet}. 
Rethinking the design of mainstream discriminators and loss functions, they tend to fully leverage the magnitude information to improve the synthesis capabilities, while neglecting the phase information, resulting in the presence of blurring artifacts. 

\begin{figure*}[!htb]
  \centering
  \includegraphics[width=0.9\textwidth]
  {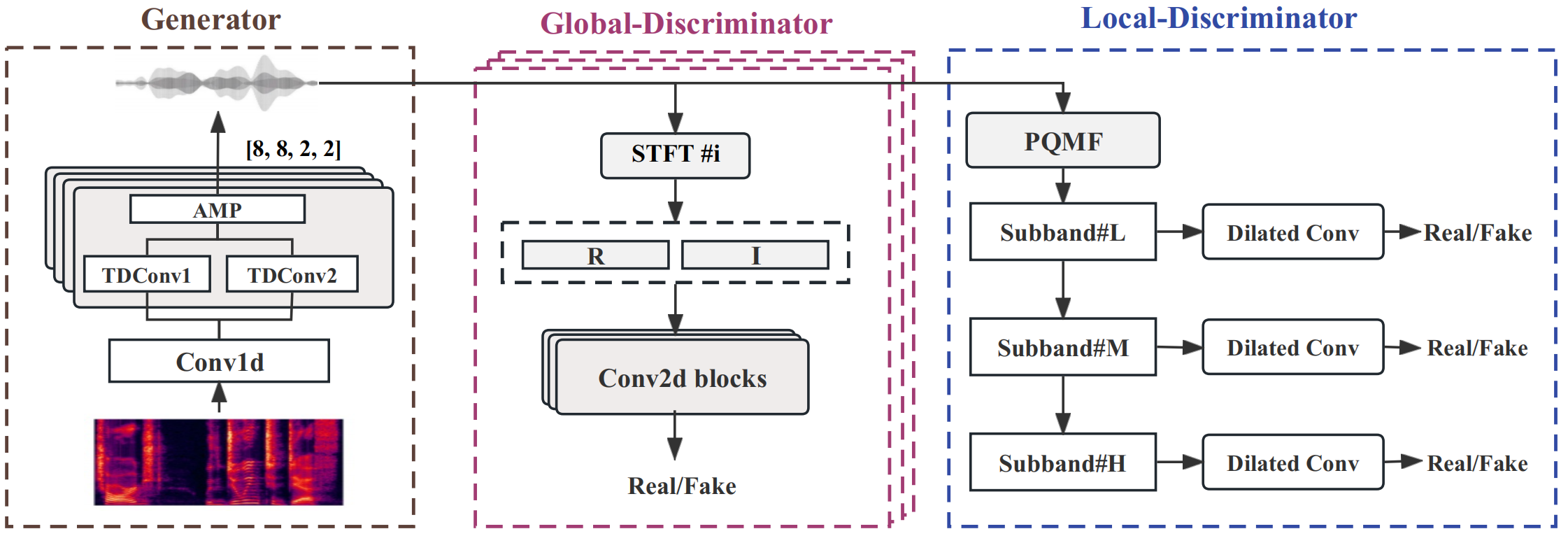}\\
  \caption{Overall architecture of FA-GAN. FA-GAN comprises an anti-aliased generator and global-level and local-level discriminators. $STFT\#i$ represents the $i$-th resolution. $\#L$, $\#M$ and $\#H$ represent low, middle and high frequency bands.}
  \label{fig:architecture}
\end{figure*}

In this paper, we propose FA-GAN, a novel generative adversarial network designed for high-fidelity speech synthesis, aiming at artifact-free and phase-aware synthesis. 
To suppress the aliasing artifacts in high-frequency areas, we improve the original structure of transposed convolution with the idea of calculating the unwanted overlap at each position. 
Furthermore, considering the difficulties of modeling phase components due to the phase wrapping issue, we shift our focus to the real and imaginary parts to supplement phase information and enhance spectral modeling abilities. 

Our contributions are summarized as follows:

\begin{itemize}
\item We analyze the main artifacts in existing vocoders, specifically upsampling artifacts and blurring artifacts, and then propose FA-GAN to synthesize high-quality speech, aiming at artifact-free and phase-aware synthesis. 
\item To alleviate the aliasing artifacts, we adopt an anti-aliased twin deconvolution module in the generator. Furthermore, we propose a novel multi-resolution RI loss to migrate the phase mismatch problem, which alleviates blurring artifacts and enriches spectral details. 
\item We evaluate both the objective and subjective performance of FA-GAN. Experimental results reveal that FA-GAN can synthesize audio samples of high fidelity and fewer artifacts and can generalize well to unseen speaker scenarios.
\end{itemize}

\section{Proposed Method}
\label{sec:method}
FA-GAN is composed of an anti-aliased generator and several discriminators as illustrated in Fig.\ref{fig:architecture}. 
The discriminators include a multi-resolution global-level discriminator and several multi-band local-level discriminators.

\subsection{Anti-aliased Generator}
\label{ssec:generator}

The backbone of the generator is inherited from HiFi-GAN~\cite{kong2020hifi} which utilizes the transposed convolution structure to upsample the resolution. 
Nonetheless, as~\cite{pons2021upsampling} revealed, this method will introduce artifacts known as checkerboard artifacts due to overlapping outputs. 
To suppress the artifacts caused by the non-ideal transposed convolution structure, we design a novel up-sampling module inspired by the work \cite{ren2023pixel}, which consists of twin deconvolution branches. 
The idea behind our proposed module is to address the issue of aliasing artifacts that emerge from the undesirable overlapping at various positions.

Specifically, we design the twin transposed convolution structure in every upsampling layer of the generator, namely, TDConv1 and TDConv2. 
The twin branch (TDConv2) is introduced in parallel with the original transposed convolution branch (TDConv1) to calculate the degree of overlap at each position. 
By performing element-by-element division between the twin branches, the unexpected artifacts of upsampling layers can be suppressed. 
Additionally, we introduce the anti-aliased multi-periodicity (AMP) block with snake activation function \cite{lee2022bigvgan,ziyin2020neural} to provide periodic inductive bias to the reconstruction process of audio, which is defined as $f(x)=x+\sin^{2}(x)$. 

\subsection{Global and Local Discriminators}
\label{ssec:discriminator}
To alleviate the blurring artifacts and enrich the spectral details, we design a multi-resolution global-level discriminator and several multi-band local-level discriminators. 

\subsubsection{Multi-resolution Global-level Discriminator}
To sharpen the structure of the full-band spectrogram, we adopt the multi-resolution complex-spectrogram discriminator as the global-level discriminator. 
Inspired by the recent progress of speech enhancement~\cite{tan2022neural}, the real and imaginary components are significant in promoting audio quality. 
To fully leverage the speech information, we utilize the stacks of the real and the imaginary spectrograms as input features of the global-level discriminator and design a novel multi-resolution RI loss function to enforce the fine-grained supervision in the frequency domain. 
Specifically, the discriminator is composed of several sub-discriminators each operating on the different real and imaginary components extracted from 2-D linear spectrograms with varying resolutions of short time fourier transform (STFT). 
We use three different scales with STFT window lengths of $[2048, 1024, 512]$ with the hop-length of $[240, 120, 50]$.

\subsubsection{Multi-band Local-level Discriminators}
Furthermore, we utilize differential pseudo quadrature mirror filter (PQMF) bank \cite{huang2022singgan,yang2021multi} to divide the full-band waveform into sub-band signals with suppressed aliasing, namely, $Subband\#L$, $Subband\#M$ and $Subband\#H$. 
Correspondingly, we design three local-level discriminators to learn various discriminative features of different sub-band signals. 
Each discriminator is composed of stacks of dilated convolutions with different dilation rates to cover diverse receptive fields. 
Through the design of local discriminators, we can leverage the discriminative features in different frequency ranges to enrich the modeling of spectral details.

\begin{figure*}[htbp]
\centering
\subfigure[Ground Truth]
{
    \begin{minipage}[b]{.15\linewidth}
        \centering
        \includegraphics[scale=0.17]{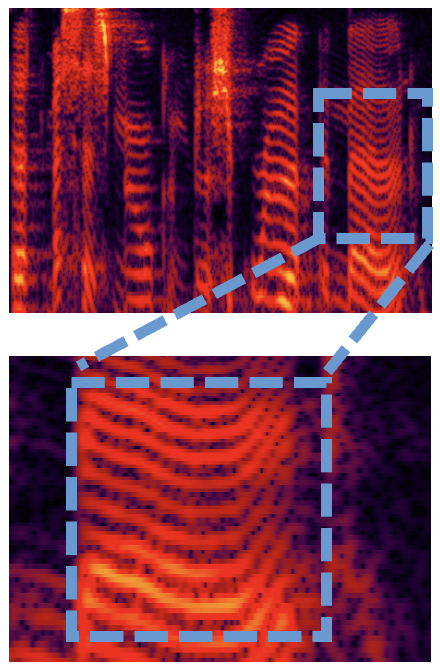}
    \end{minipage}
}
\subfigure[HiFi-GAN]
{
 	\begin{minipage}[b]{.15\linewidth}
        \centering
        \includegraphics[scale=0.17]{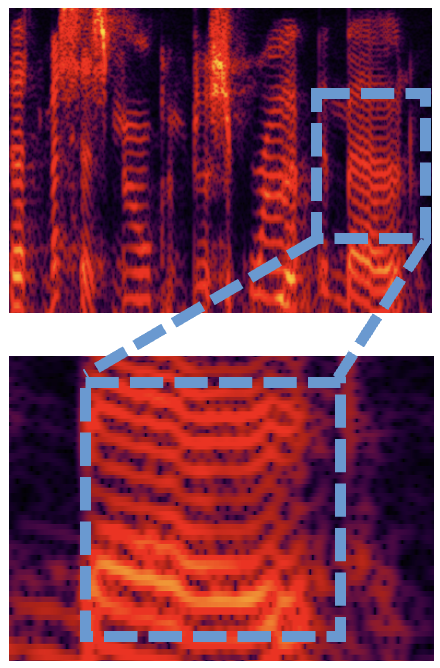}
    \end{minipage}
}
\subfigure[UnivNet-c32]
{
 	\begin{minipage}[b]{.15\linewidth}
        \centering
        \includegraphics[scale=0.17]{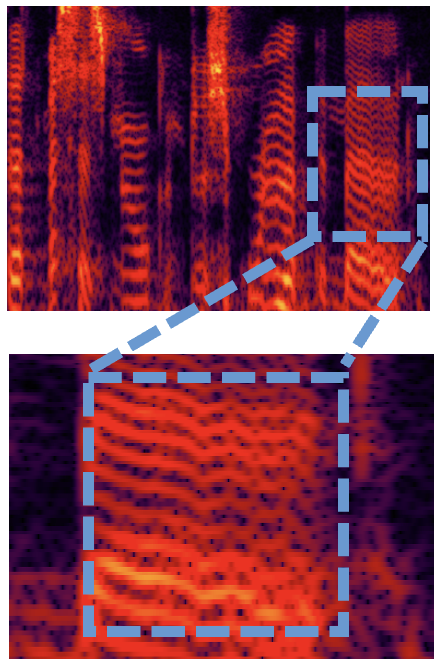}
    \end{minipage}
}
\subfigure[Avocodo]
{
 	\begin{minipage}[b]{.15\linewidth}
        \centering
        \includegraphics[scale=0.17]{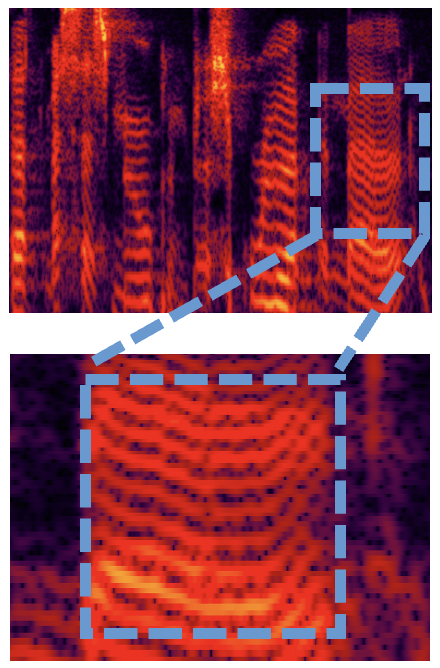}
    \end{minipage}
}
\subfigure[BigVGAN]
{
 	\begin{minipage}[b]{.15\linewidth}
        \centering
        \includegraphics[scale=0.17]{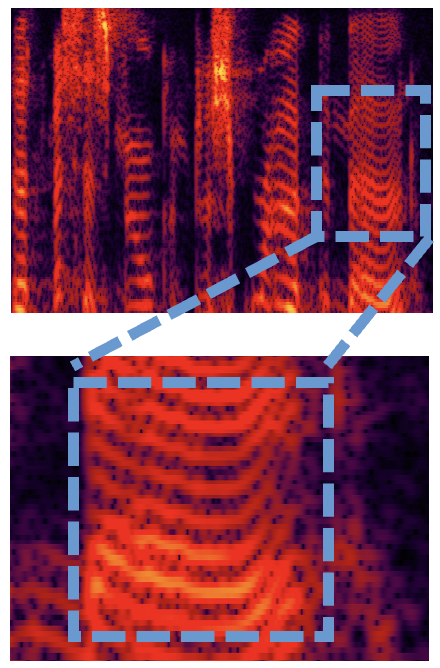}
    \end{minipage}
}
\subfigure[FA-GAN]
{
 	\begin{minipage}[b]{.15\linewidth}
        \centering
        \includegraphics[scale=0.17]{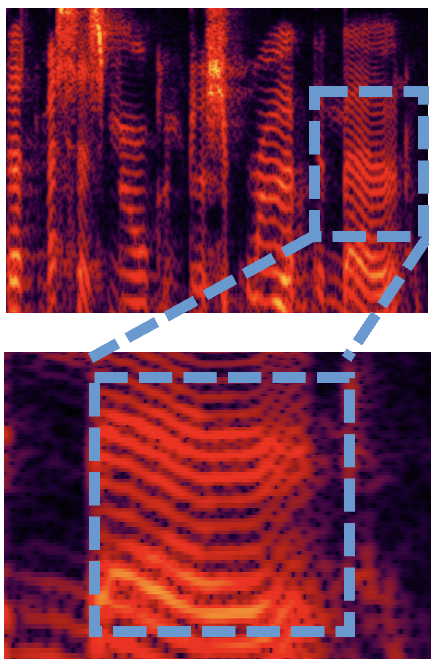}
    \end{minipage}
}
\caption{Visualization of spectrograms generated from FA-GAN and baseline vocoders. The bottom row offers an enlarged perspective of high-frequency components to illustrate spectral differences.}
\label{fig:visual}
\end{figure*}

\subsection{Training Objectives}
\label{ssec:loss}
The training loss is composed of multi-resolution RI loss, adversarial loss, mel loss, and feature matching loss.

\subsubsection{Multi-resolution RI Loss} 
In reevaluating the architecture of prevalent discriminators and loss functions, it can be observed that these frameworks tend to leverage magnitude information to enhance synthesis quality, yet often overlook the significance of phase information~\cite{kong2020hifi,jang2021univnet,bak2023avocodo}. 
It is well known that speech signals can be decomposed into real and imaginary components via STFT. 
During the training of vocoders, phase information is implicitly reconstructed by the generator, making phase mismatch a significant problem in vocoder modeling.
Given the challenges in modeling phase features, especially due to phase wrapping issues, our focus shifts towards the real and imaginary components. 
We propose a novel loss function to enforce the alignment of the real and imaginary parts of the real audio $x$ and its reconstruction $\hat{x}$. 
This method fully utilizes the frequency domain decomposition ability of STFT, providing richer and more precise frequency domain information. 
We extract the real and imaginary parts of $x$ and $\hat x$ by STFT and enforce the frequency spectral regularization through $L_{1}$ norm.
The loss is defined as follows:
\begin{equation}
\begin{split}
\{R_x, I_x \} \leftarrow STFT(x),~
\{ \hat{R_x}, \hat{I_x}\} \leftarrow STFT(\hat{x}), 
\end{split}
\end{equation}
\begin{equation}
\begin{split}
L_{RI}(x, \hat x) &= | \hat{R_x}-R_x |_{1} + | \hat{I_x}-I_x |_{1} \\ 
& + | \sqrt{{\hat{R_x}}^{2} +\hat{I_x}^{2}} - |STFT(x)||_{1} \\
& +\frac{|STFT(x)-STFT(\hat x)|_{F} }{|STFT(x)|_{F} }, 
\end{split}
\end{equation}
where $STFT( \cdot)$ denotes the short-time Fourier transform and extracts a complex spectrogram, as well as $R$ and $I$ represent the real and imaginary parts of audio samples respectively. 

To model different scales of frequency information better, we extend the RI loss to the multi-resolution one with different analysis parameters (i.e., FFT size, frame shift, and window size). 
The multi-resolution RI loss is defined as follows: 
\begin{equation}
\begin{split}
L_{MR-RI}(x, \hat x) = \frac{1}{M} \sum_{m=1}^{M}L_{RI}^{(m)}(x,\hat x). 
\end{split}
\end{equation}

\subsubsection{Adversarial Loss} 
GAN losses for the generator $G$ and the discriminator $D$ are defined as follows:
\begin{equation*}
L_{adv}(G;D_{n})=E_{(x,s)}\left [ (1-D_{n}(x_{n}))^{2} + (D_{n}(y_{n}))^{2}    \right ], 
\end{equation*}
\begin{equation}
L_{adv}(D_{n};G) = E_{(s)} \left [ \sum_{n=1}^{N} (1-D_n(y_{n}))^{2} \right ],
\end{equation}
where $x$ and $y$ denote the full-band ground truth and the generated audio sample, while $x_{n}$ and $y_{n}$ denote the $n$-th sub-band audio signal of ground truth and generated samples, respectively. 
$s$ denotes the mel-spectrogram of ground truth. 
$D_{n}$ represents the $n$-th discriminator and $N$ represents the number of discriminators.

\subsubsection{Final Loss} 
Our approach is optimized by the following objective function with the above losses. 
Especially, $L_{mel}$ and $L_{fm}$ denote the mel loss and feature matching loss as HiFi-GAN \cite{kong2020hifi}. 
\begin{equation}
\begin{split}
L_{G}&=\lambda_{g}\sum_{n=1}^{N}L_{adv}(G;D_{n})+\lambda_{RI}L_{MR-RI} \\
&+\lambda_{mel}L_{mel}+\lambda _{fm}L_{fm}, 
\end{split}
\end{equation}
\begin{equation}
\begin{split}
L_{D}=\sum_{n=1}^{N} L_{adv}(D_{n};G),
\end{split}
\end{equation}
where $\lambda_{g}$, $\lambda_{RI}$, $\lambda_{mel}$, and $\lambda_{fm}$ are the scalar coefficients to balance between the loss terms.

\begin{table*}[]
\centering
\caption{Objective evaluation results for the seen speaker and unseen speaker scenarios and metrics are MCD, $F_{0}$-RMSE, PESQ, and LSD, respectively. For PESQ, higher scores indicate better performance, whereas for other metrics, lower scores are preferable. The best results for each metric are highlighted in bold. }
\label{tab:obj results}
\begin{tabular}{ccccc|cccc}
\hline
& \multicolumn{4}{c|}{ \textbf{Seen speaker (LJSpeech)}}  & \multicolumn{4}{c}{ \textbf{Unseen speaker (VCTK)}} \\ \hline
\multicolumn{1}{c|}{\textbf{Model}}    & \textbf{MCD ($\downarrow$)}    & \textbf{$F_{0}$-RMSE ($\downarrow$)} & \textbf{LSD ($\downarrow$)}   & \textbf{PESQ ($\uparrow$)}    & \textbf{MCD ($\downarrow$)}    & \textbf{$F_{0}$-RMSE ($\downarrow$)} & \textbf{LSD ($\downarrow$)}   & \textbf{PESQ ($\uparrow$)}    \\ \hline
\multicolumn{1}{c|}{HiFi-GAN~\cite{kong2020hifi}}          & 2.6843 & 34.6936 & 0.8456 & 3.4093 & 4.1821 & 42.9041 & 1.0344 & 2.3259 \\
\multicolumn{1}{c|}{UnivNet-$c32$~\cite{jang2021univnet}}  & 2.6325 & 35.5299 & 0.8523 & 3.5317 & 4.1070 & 42.7174 & 1.0218 & 2.2433  \\
\multicolumn{1}{c|}{Avocodo~\cite{bak2023avocodo}}         & 2.5855 & 35.2415 & 0.8212 & 3.6488 & 3.9296 & 41.3385 & 1.0291 & 2.2671 \\
\multicolumn{1}{c|}{BigVGAN~\cite{lee2022bigvgan}}         & 2.5824 & 34.5967 & 0.8201 & 3.6781 & 3.9017 & 41.7834 & 0.9833 & 2.3152 \\ \hline
\multicolumn{1}{c|}{FA-GAN}                                & 2.5796 & 34.5718 & 0.8079 & 3.6862 & 3.8756 & 39.7165 & 0.9379 & 2.4286 \\ \hline
\multicolumn{1}{c|}{FA-GAN+AUG}                            & \textbf{2.5447} & \textbf{34.4667} & \textbf{0.7724} & \textbf{3.7019} & \textbf{3.6912} & \textbf{38.3704} & \textbf{0.9235} & \textbf{2.5328} \\ \hline
\end{tabular}
\end{table*}

\begin{table}[!t]
\centering
\setlength{\tabcolsep}{1.4mm}
\caption{Subjective evaluation results in terms of 5-scale MOS with 95\% confidence intervals.}
\label{tab:MOS}
\begin{tabular}{ccccc}
\hline
\multirow{2}{*}{\textbf{MOS}} & \multicolumn{2}{c}{\textbf{Seen speaker}} & \multicolumn{2}{c}{\textbf{Unseen speaker}} \\ \cline{2-5} 
         & \textbf{MOS ($\uparrow$)}      & $95$\% CI   & \textbf{MOS ($\uparrow$)}     & $95$\% CI \\ \hline
Ground Truth                            & 4.432    & 0.08   & 4.518   & 0.05      \\ \hline
HiFi-GAN~\cite{kong2020hifi}            & 3.931    & 0.08   & 3.872   & 0.07      \\ 
UnivNet-$c32$~\cite{jang2021univnet}    & 3.819    & 0.09   & 3.714   & 0.09      \\
Avocodo~\cite{bak2023avocodo}           & 4.025    & 0.08   & 3.893   & 0.07      \\
BigVGAN~\cite{lee2022bigvgan}           & 4.137    & 0.08   & 3.928   & 0.07      \\ \hline     
\multicolumn{1}{c}{FA-GAN}              & 4.193    & 0.07   & 3.973   & 0.07      \\ \hline
\multicolumn{1}{c}{FA-GAN+AUG}          & \textbf{4.215}    & \textbf{0.06}    & \textbf{4.182}    & \textbf{0.07}   \\ \hline
\end{tabular}
\end{table}

\section{Experiments}
\label{sec:exp}

\subsection{Experimental Setups}
\label{ssec:exp}

We conduct experiments on LJSpeech~\cite{ljspeech17} and VCTK~\cite{vctk19}. 
For the LJSpeech dataset, we randomly divide the dataset into the training set, validation set, and test set, $80$\%, $10$\%, and $10$\% respectively. 
For the VCTK dataset, we randomly select the audio samples of $10$ speakers as the unseen speaker test set. 
All audio samples are downsampled to $22050$ Hz. 

Moreover, $80$-dimensional mel-spectrograms are used as input features, which are calculated with the short-time Fourier transform. 
The FFT, window, and hop size are set to $1024$, $1024$, and $256$, respectively.

Four popular GAN-based models are selected to be compared with FA-GAN, including HiFi-GAN V1, UnivNet-$32c$, Avocodo and BigVGAN. 
Especially, they were all trained up to $1M$ steps for equal comparison and the hyper-parameters of FA-GAN are the same as those of HiFi-GAN.

\subsection{Data Augmentation Strategies}
To improve the generalization ability of the vocoder with limited data, we perform several data augmentation tricks on the training dataset to simulate unseen speaker scenarios. 

\textbf{Harmonic Shift.}
To improve the model's generalization ability for unseen speakers, we utilize parselmouth \cite{jadoul2018introducing} to modify the $F_{0}$ and formant to achieve harmonic shift, imitating the timbre of different speakers. 

\textbf{Lossy Compression.}
To improve the robustness of vocoder, we use $Opus$ as the lossy compression codec method to encode the real audio with target bitrates of $32$ kbps. 

\textbf{Global Noise.}
We add random noise on the original data with the range of [$28$dB, $40$dB] to simulate the audio data in the real scenarios.

\subsection{Evaluation Metrics} 
We conduct both objective evaluations and subjective evaluations. 
For objective evaluation, we calculate the mel-cepstral distortion (MCD) \cite{kubichek1993mel} and perceptual evaluation of speech quality (PESQ) \cite{rix2001perceptual} to evaluate the quality of the audio. 
Moreover, we measure the $F_{0}$ root mean square error ($F_{0}$-RMSE) to evaluate the reproducibility of $F_{0}$ and the log-spectral distance (LSD) to measure the spectral differences between the ground truth and generated samples. 

For the subjective evaluation, we conduct 5-scale mean opinion score (MOS) tests to evaluate the quality of synthesized audios. 
Specifically, we invited 10 participants to score the sound quality of 120 audio samples. 

\subsection{Results and Analysis}
\label{ssec:obj}
We perform both objective evaluations and subjective evaluations under the seen and unseen scenarios to compare our proposed FA-GAN with other popular vocoders quantitatively in terms of audio quality and artifact suppression. 

\subsubsection{Audio Quality \& Comparison} 
As shown in Table~\ref{tab:obj results} and Table~\ref{tab:MOS}, FA-GAN outperforms other vocoders on both objective and subjective metrics. 
Specifically, FA-GAN outperforms the baseline models in terms of objective MCD and PESQ scores and subjective MOS scores, which reveals that FA-GAN has better audio quality. 
Moreover, FA-GAN achieves lower $F_{0}$-RMSE and LSD score for that we perform fine-grained supervision on the frequency domain through multi-resolution RI loss. 
We further reveal that FA-GAN can well generalize to unseen speaker scenarios. 
Moreover, leveraging various data augmentation strategies mentioned above to enrich the diversity of fake data, denoted as FA-GAN+AUG, can further improve performance.

\subsubsection{Artifacts Visualization} 
In Fig.~\ref{fig:visual}, we make a visualization of the spectrograms generated by the ground truth and other vocoders, such as HiFi-GAN, UnivNet-$c32$, Avocodo, and BigVGAN. 
It can be observed that mel-spectrograms generated by other vocoders suffer from noticeable artifacts and the harmonic details are aliasing and blurring. 
By contrast, FA-GAN has more explicit high-frequency harmonic details for the reason of fine-grained supervision on the frequency domain.

\subsubsection{Ablation Studies}
\label{ssec:ablation}
In Table~\ref{tab:ablation}, we conduct ablation studies to observe the effectiveness of each component in FA-GAN. 
We observe that all objective metrics, including MCD, $F_{0}$-RMSE, and LSD, decline when the twin deconvolution (TDConv) module is replaced with the transposed convolution. 
The main reason is that the original transposed convolution brings aliasing artifacts, leading to audio quality degradation. 
Additionally, a sharp decline in model performance is observed when the multi-resolution real and imaginary (RI) loss is removed, demonstrating that the real and imaginary parts are of great significance in improving audio quality and suppressing spectral artifacts. 
Thus, it can be concluded that both the artifacts caused by non-ideal transposed convolution and the lack of phase information significantly affect the synthesized speech quality, and these issues should be taken seriously.

\begin{table}[!t]
\centering
\caption{Ablation results in the seen speaker scenario.}
\label{tab:ablation}
\begin{tabular}{ccccc}
\hline
\textbf{Model}           & \textbf{MCD ($\downarrow$)}   & \textbf{$F_{0}$-RMSE ($\downarrow$)}   & \textbf{LSD ($\downarrow$)}   \\ \hline
FA-GAN        & 2.580  & 34.572     & 0.808   \\
w/o TDConv      & 2.593  & 34.618     & 0.835   \\ 
w/o MR-RI loss     & 2.749  & 35.893     & 0.862   \\ \hline
\end{tabular}
\end{table}

\section{Conclusion}
\label{conclusion}
In this paper, we propose FA-GAN, a novel GAN-based vocoder for high-fidelity speech synthesis with suppressed artifacts. 
Considering the aliasing artifacts caused by imperfect upsampling layers, we introduce the twin deconvolution structure to suppress artifacts in high-frequency areas. 
Moreover, we fully leverage the complex-valued spectrograms and design a novel loss in terms of the real and imaginary components to perform fine-grained supervision on the frequency domain, which alleviates the blurring artifacts and enriches the spectral details. 
Various comparative experimental results reveal that FA-GAN has achieved impressive performances in both audio quality and artifact suppression.

\section{Acknowledgements}
\label{ack}
This work is supported by the Natural Science Foundation of China (NSFC) under the grant NO. 62172306, Hubei Province Technological Innovation Major Project (NO. 2021BAA034, 2020BAB018).



\bibliographystyle{IEEEtran}
\bibliography{mybib}

\end{document}